\documentclass[conference]{IEEEtran}
\IEEEoverridecommandlockouts

\usepackage{graphicx}
\usepackage{subfigure}
\usepackage{float}
\usepackage{amsmath,amssymb,amsfonts}
\usepackage{textcomp}
\usepackage{makecell}
\usepackage{cite}
\usepackage{xcolor}
\usepackage[english]{babel}
\usepackage{bbm}
\usepackage{soul}
\usepackage[nolist]{acronym}
\usepackage{multirow}
\usepackage{multicol}
\usepackage{tabularx}
\usepackage{pifont}
\usepackage{booktabs}
\usepackage{tikz}
\usetikzlibrary{positioning,arrows.meta}
\usepackage{algpseudocode}
\usepackage[linesnumbered, lined, ruled, noend]{algorithm2e}
\usepackage[flushleft]{threeparttable}

\def\BibTeX{{\rm B\kern-.05em{\sc i\kern-.025em b}\kern-.08em
    T\kern-.1667em\lower.7ex\hbox{E}\kern-.125emX}}

\begin{acronym}
\acro{LLM}{Large Language Model}
\acro{BS}{Base Station}
\acro{ABS}{Aerial Base Station}
\acro{UAV}{Unmanned Aerial Vehicle}
\acro{AI}{Artificial Intelligence}
\acro{LSTM}{Long Short-Term Memory}
\acro{CNN}{Convolutional Neural Network}
\acro{QoS}{Quality of Service}
\acro{QoE}{Quality of Experience}
\acro{RAG}{Retrieval-Augmented generation}
\acro{RSU}{Roadside Unit}
\acro{MAC}{Multiple-Access Control}
\acro{E2E}{End-to-End}
\acro{UE}{User Equipment}
\acro{LoS}{Line-of-Sight}
\acro{PoA}{Point of Attachment}
\acro{MINLP}{Mixed Integer Non-Linear Programming}
\acro{DRL}{Deep Reinforcement Learning}
\acro{DNN}{Deep Neural Network}
\acro{D3QL}{Dueling Double Deep Q-Learning}
\acro{GNN}{Graph Neural Network}
\acro{NFV}{Network Function Virtualization}
\acro{VNF}{Virtual Network Function}
\acro{B5G}{Beyond 5G}
\acro{SDN}{Software-Defined Network}
\acro{AR}{Augmented Reality}
\acro{VR}{Virtual Reality}
\acro{XR}{Extended Reality}
\acro{SLA}{Service-Level Agreement}
\acro{6G}{sixth-generation}
\acro{MEC}{Mobile Edge Cloud}
\acro{RIS}{Reconfigurable intelligent surface}

\acro{Mbps}{Megabit per second}
\acro{Gbps}{Gigabit per second}
\acro{Tbps}{Terabit per second}
\acro{GFLOPS}{giga-floating point operations per second}
\acro{SUMO}{Simulation of Urban Mobility}

\acro{QoAIS}{Quality-of-AI-Services}
\acro{HyPE}{\underline{hy}brid \underline{p}redictive-in-context-l\underline{e}arning}
\acro{MAP}{\underline{m}obility-\underline{a}ware \underline{p}rediction}
\acro{LEAD}{\underline{le}arning-\underline{a}ugmented \underline{d}ecision}
\acro{SET}{\underline{s}ervice plac\underline{e}ment and rou\underline{t}ing}
\acro{LBSP}{Latency-Biased Shortest Path}
\acro{PGA}{Proximity-Greedy Allocation}
\acro{AD-SAC}{Action-Decoupled Soft Actor-Critic}
\acro{JAAPD-D}{Joint AI Agent Placement with Deep neural network Deployment}

\end{acronym}

\DeclareFontFamily{U}{mathx}{}
\DeclareFontShape{U}{mathx}{m}{n}{<-> mathx10}{}
\DeclareSymbolFont{mathx}{U}{mathx}{m}{n}
\DeclareMathAccent{\widehat}{0}{mathx}{"70}
\DeclareMathAccent{\widecheck}{0}{mathx}{"71}
\DeclareMathAlphabet{\mathpzc}{OT1}{pzc}{m}{it}

\usepackage{fancyhdr, lipsum}
\fancypagestyle{mahmood}{%
   \fancyhf{} 
   
   \fancyhead[C]{\copyright~2026 IEEE. Reprinting or republishing this material for the purpose of advertising or promotion, creating new collective works, reselling or redistributing to servers or lists, or using any copyrighted component in other works must adhere to IEEE policy. The paper has been accepted for publication at \textbf{IEEE FINE 2026}.}
}%

\makeatletter
\let\ps@IEEEtitlepagestyle\ps@mahmood
\makeatother

\begin{document}

\title{Quality-Aware Personalized AI Service Provisioning in UAV-Assisted 6G Networks}


\author{
    \IEEEauthorblockN{
        Mohammad Farhoudi\textsuperscript{1}, Masoud Shokrnezhad\textsuperscript{2}, and Tarik Taleb\textsuperscript{3}\\
    }
    \IEEEauthorblockA{
        \textsuperscript{1} \textit{Oulu University, Finland}; mohammad.farhoudi@oulu.fi \\
        \textsuperscript{2} \textit{ICTFICIAL Oy, Espoo, Finland}; masoud.shokrnezhad@ictficial.com \\
        \textsuperscript{3} \textit{Ruhr University Bochum (RUB), Germany}; tarik.taleb@rub.de
    }
}

\maketitle

\begin{abstract}
In sixth-generation (6G) artificial intelligence (AI) services, two quality dimensions should be jointly addressed: conventional quality (e.g., latency) and Quality of AI Services (QoAIS; output fidelity, continuity, personalization). Existing methods emphasize conventional quality, while neglecting QoAIS, particularly for personalized outputs in dynamic aerial-terrestrial settings. This paper introduces HyPE, a \underline{hy}brid \underline{p}redictive-in-context-l\underline{e}arning framework for holistically quality-aware personalized AI service provisioning in Unmanned Aerial Vehicle (UAV)-assisted 6G networks. HyPE integrates: (i) mobility-aware prediction to forecast spatio-temporal request distributions, (ii) learning-augmented decision leveraging Large Language Model (LLM)-based reasoning to optimize UAV trajectories and inference assignments, and (iii) pre-/post-processing service placement and routing using heuristics. We formulate an optimization problem for joint trajectory planning, service placement, and routing, and present HyPE as a scalable alternative to intractable optimal solutions. Simulations with empirical mobility traces and heterogeneous AI workloads show near-optimal coverage, reduced end-to-end latency, sustained QoAIS-driven, and continuity-based service personalization versus optimization and state-of-the-art baselines. The results highlight the promise of predictive learning-augmented provisioning for elastic, user-centric AI in 6G.
\end{abstract}

\begin{IEEEkeywords}
Service Provisioning, Quality of AI Services (QoAIS), Edge-Cloud Environment, Intelligent UAV, 6G Aerial-Terrestrial Networks.
\end{IEEEkeywords}

\vspace{-4pt}
\section{Introduction}
With the rising demand for \ac{AI}-driven services, \ac{6G} networks should pave the way to accommodate them. \Ac{6G} is envisioned as an \ac{AI}-native infrastructure, enabling adaptive service provisioning \cite{farhoudi2025Survey, mazandaraniGlobecom2025}. 
It should jointly satisfy two service quality dimensions: conventional quality (\ac{E2E} latency) and \ac{QoAIS} (output fidelity and continuity-aided user-specific personalization) \cite{gao2025, AIQuality2023}. Meeting these dual requirements calls for a heterogeneous model stack, where edge-cloud nodes host lightweight distilled generative models for low-latency responses alongside full-scale models for richer reasoning and \ac{QoAIS}-driven personalization that adapts outputs to user history \cite{zhang2025surveygraphretrievalaugmentedgeneration, chen2025surveycollaborative}. To remain scalable under high load, this satisfaction should be modular, leveraging pluggable pre-processing (e.g., filtering, visual encoding) and post-processing (e.g., voice modulation, translation, formatting) stages that can be composed per request to balance \ac{E2E} latency and \ac{QoAIS} \cite{hao2025rapretrievalaugmentedpersonalizationmultimodal}. Additionally, resource-intensive inference pipelines should be orchestrated in dynamic networks, posing a central challenge for efficient provisioning under variable scenarios \cite{wu2024personalizedmultimodallargelanguage}.

Furthermore, mobility adds complexity to service provisioning, as users expect reliable responses while moving across network regions \cite{farhoudi2024}. \Ac{6G} networks aim to provide ubiquitous access through \ac{AI}-native infrastructure and intelligent edge capabilities \cite{6garchitecture_tarik}. To realize such pervasive access, \ac{6G} network architectures incorporate aerial platforms as mobile edge nodes that complement the static terrestrial infrastructure \cite{mazandarani2025}. Equipped with computing resources and embedded \ac{AI} models, \acp{UAV} deliver on-demand coverage and adaptive service provisioning, executing pre-/post-processing and inference tasks. Dynamic transitions of these tasks are therefore essential to maintain \ac{QoAIS} across user sessions in highly mobile environments.

Recent research has investigated task offloading, service placement, and pipeline scheduling in edge-cloud environments under latency and accuracy constraints. For instance, Hao \textit{et al.} \cite{multiUAVoffloading} proposed a discrete-continuous \ac{DRL} method based on latent space to optimize \ac{UAV} trajectories and resource allocation for latency reduction, while Ding \textit{et al.} \cite{AIQuality2023} introduced a multi-objective scheduler, using genetic-based algorithms for \ac{QoAIS}-aware provisioning to balance latency and accuracy. Raj \textit{et al.} \cite{Raj2025} focused on personalized offloading by deadline-aware and guaranteed experience heuristic methods to assist visually impaired users, whereas Jin \textit{et al.} \cite{Offloading2025} proposed multi-tier architectures and game-theoretic offloading strategies that dynamically allocate resources to minimize latency under vehicular mobility. Yan \textit{et al.} \cite{Yan2025} presented \ac{AD-SAC} for accuracy-aware \ac{LLM} offloading in \ac{UAV}-satellite networks, jointly optimizing trajectory and model placement to minimize total service latency. Similarly, Hu \textit{et al.} \cite{Hu2024} developed \ac{JAAPD-D}, combining multi-timescale Lyapunov optimization with \ac{AI} model placement to balance latency and acceptance rates.

Despite these advances, several important gaps remain.
Existing studies mainly fall into two separate lines: they either address generic \ac{AI} task offloading, resource allocation, or service placement without explicitly supporting personalized inference, or they consider personalized \ac{AI}-oriented services without jointly optimizing \ac{UAV} trajectory control, inference node assignment, and service continuity under mobility. Current solutions mostly overlook \ac{QoAIS}, especially the joint interplay among \ac{E2E} latency, model fidelity, and continuity-aware personalization in dynamic aerial-terrestrial \ac{6G} networks. In particular, limited attention has been paid to the spatio-temporal alignment between \ac{UAV} movement and the availability of user-specific historical context required for personalized inference, as well as to the coordinated use of distilled and full-scale models across heterogeneous nodes to satisfy different latency-quality demands.
These limitations call for an adaptive network-layer enabler that proactively tracks user mobility and requests; positions \acp{UAV} near users with relevant personalization history; assigns appropriate model variants according to latency and quality 
requirements; and distributes pre-processing, inference, and post-processing functions across heterogeneous aerial-terrestrial resources. 
In response, this paper introduces a mobility-aware, latency- and \ac{QoAIS}-driven framework for personalized \ac{AI} service provisioning with four key contributions:
\begin{itemize}
\item a modular service fabric that composes pluggable pre- and post-processing stages to elastically serve requests, enabling cross-modal latency and \ac{QoAIS} co-optimization with on-the-fly model distillation; 

\item a joint optimization formulation that couples user mobility for \ac{UAV} trajectory planning, function placement, inference assignment, and routing under quality and resource constraints, balancing \ac{E2E} latency, output fidelity, and personalization continuity;

\item a hybrid predictive-learning stack that fuses \ac{DRL}-based spatio-temporal demand forecasting with structured \ac{LLM}-guided online orchestration, and lightweight heuristics for constraint-aware routing and placement; and 

\item extensive evaluations on mobility traces and heterogeneous \ac{AI} workloads showing near-optimal coverage, consistently low latency, and sustained \ac{QoAIS}-driven personalization and output fidelity under high user density.
\end{itemize}

The remainder of the paper is organized as follows: Section \ref{sec:system_model} presents the system model, including network architecture and service pipeline; Section \ref{sec:problem} formulates the problem; Section \ref{sec:method} details the proposed hybrid framework; Section \ref{sec:results} provides performance evaluation; and Section \ref{sec:conclusion} concludes and outlines future research directions.

\section{System Model}\label{sec:system_model}
In this section, we detail a system designed to deliver \ac{AI} services to mobile users over dynamic \ac{6G} networks.

\subsection{Network Architecture}
We model the system as a grid $\boldsymbol{\mathcal{A}}$ of discrete areas reflecting urban sectors or road segments. Area adjacency is captured by the binary predicate $\mathcal{A}_{a,a'}$, set to 1 if $a$ and $a'$ are neighbors and 0 otherwise. The network is modeled as a time-varying directed graph $\mathcal{G}(\mathcal{N}, \mathcal{L}, \mathcal{P})_t$, where $\boldsymbol{\mathcal{N}}$, $\boldsymbol{\mathcal{L}}_t$, and $\boldsymbol{\mathcal{P}}_t$ represent the set of network nodes, links, and candidate routing paths at time $t \! \in \! \boldsymbol{\mathcal{T}}$. The multi-tier network comprises cloud nodes, terrestrial edge servers (e.g., roadside units or base stations), and aerial \ac{UAV} computing nodes. Each node $n \in \boldsymbol{\mathcal{N}}$ is characterized by its type, computing capacity $\widehat{\mathcal{C}}_n$, and model fidelity score $\overline{Q}_n$, capturing the quality of the \ac{AI} model hosted at $n$. Communication links $l \in \boldsymbol{\mathcal{L}}_t$ describe feasible wired/wireless connections among nodes, characterized by capacity $\widehat{\mathcal{L}}_l$ and determined by network topology and aerial node positions. Depending on available links at each time frame, a set of paths $\boldsymbol{\mathcal{P}}_t$ defines feasible packet transmission routes, with $\mathcal{H}_{p,t}$ as well as $\mathcal{T}_{p,t}$ denoting head and tail nodes, and $\mathcal{J}_{p,l,t}$ indicating whether path $p$ traverses link $l$ at time $t$. Time is slotted, and each frame is long enough for one adjacency-constrained UAV move and service reconfiguration; thus, UAV movement time is absorbed into frame evolution.

\vspace{-3pt}
\subsection{AI Services}
The system supports a catalog $\boldsymbol{\mathcal{S}}$ of \ac{AI} services (e.g., voice command processing). Each service $s$ with duration $\mathcal{T}^{du}_{s}$ is structured as a three-stage chain: (i) \textit{Pre-processing} $\boldsymbol{\mathcal{F}}^{pr}_s$: lightweight functions with $\pi^{pr}_s$ stages that prepare input for inference; (ii) \textit{Inference}: reasoning using one \ac{AI} model, with full generative models on edge-cloud nodes or distilled variants on \acp{UAV}; and (iii) \textit{Post-processing $\boldsymbol{\mathcal{F}}^{po}_s$}: final transformations before serving the user, with $\pi^{po}_s$ stages. This separation enables the selection of serving nodes with different fidelity levels under compute, communication, and latency constraints.

\subsection{User and Request}
Users $u \!\in\! \boldsymbol{\mathcal{U}}$ generate requests $r \in \boldsymbol{\mathcal{R}}_u$ - in which $\boldsymbol{\mathcal{R}}_u \!=\! \{r \!\in\! \boldsymbol{\mathcal{R}} ~\big|~ r$ is generated by user $u \}$ - over time horizon $\boldsymbol{\mathcal{T}}$, connecting through their \ac{PoA} to access the required services.
Each request is represented by a tuple ($\mathcal{S}_r, \mathcal{T}^{pr}_r, \Delta_r, \mathcal{I}_{u_r,a,t}, \widecheck{\mathcal{Q}}^{ou}_{r}, \widecheck{\mathcal{Q}}^{la}_{r}$), specifying the requested service, start time, admissible request window $\Delta_r\!=\![\mathcal{T}^{pr}_r,\mathcal{T}^{pr}_r\!+\!\mathcal{T}^{du}_{s_r}]$, user mobility path, minimum acceptable inference fidelity, and maximum tolerable \ac{E2E} latency, respectively. Successful service delivery requires meeting both $\widecheck{\mathcal{Q}}^{ou}_{r}$ and $\widecheck{\mathcal{Q}}^{la}_{r}$.
Additionally, personalization quality $\mathcal{Q}^{pe}_{u}$ is treated as a continuity proxy: repeated service at previously associated inference nodes improves context retention and response consistency. Each request $r$ also requires minimum bandwidth $\widecheck{\mathcal{L}}_r$ along with maximum packet size $\widecheck{\mathcal{Z}}_{r,t}$, influencing transmission feasibility and latency, with latency modeled as path-based aggregation. Finally, requests' computational demands are defined in \ac{GFLOPS} as $\widecheck{\mathcal{C}}^{pr}_{r,f},\widecheck{\mathcal{C}}^{in}_{r},\widecheck{\mathcal{C}}^{po}_{r,f}$ for pre-, inference, and post-processing, respectively.


\section{Problem Formulation}\label{sec:problem}
The optimization framework aims to jointly maximize the number of supported requests and their \ac{QoAIS} while minimizing \ac{E2E} latency for \ac{AI} services in \ac{6G} networks. The objective function (OF) balances three terms: output fidelity weighted by $\alpha$ (reflecting node suitability), personalization quality scaled by $\beta$ (promoting assignment continuity), and latency penalized by $\zeta$ (aggregated over routing paths). Here, $\alpha$, $\beta$, and $\zeta$ are weighting coefficients that tune the relative importance of fidelity, personalization, and latency despite their different numerical scales. Request acceptance is represented by the binary variable $\mathcal{X}_{r}$, with respect to $\mathcal{X}^{pr}_{r,f,t},\mathcal{X}^{in}_{r,t},\mathcal{X}^{po}_{r,f,t}$ denoting execution of pre-processing, inference, and post-processing functions, respectively.
To unify notation across service stages, we use the stage-dependent variable $\mathcal{X}^{\phi}_{r,f,t}$. For pre- and post-processing stages, $f$ indexes functions in $F^{\phi}_{s}$, while for inference, we define a singleton set $\boldsymbol{\mathcal{F}}^{in}_s=\{\varnothing\}$, so that $\mathcal{X}^{in}_{r,f,t}=\mathcal{X}^{in}_{r,t}$.
A request is considered accepted if its three service stages are completed within the prescribed temporal structure: all pre-processing functions are executed at the entry frame $\mathcal{T}^{pr}_r$, inference is executed during $\mathcal{T}^{in}_r\!=\!(\mathcal{T}^{pr}_r, \mathcal{T}^{pr}_r \!+\! \mathcal{T}^{du}_{s_r})$, and all post-processing functions are completed at the terminal frame $\mathcal{T}^{po}_r = \mathcal{T}^{pr}_r \!+\! \mathcal{T}^{du}_{s_r}$ (C1).

\footnotesize
\begin{align}\label{OBJECTIVEFUNCTION1}
& \mathrm{ max } \sum_{\boldsymbol{\mathcal{R}}} \! \mathcal{X}_r + \alpha \! \cdot \! \sum_{\boldsymbol{\mathcal{R}}} \! \mathcal{Q}^{ou}_{r} \! + \! \beta \! \cdot \! \sum_{\boldsymbol{\mathcal{U}}} \! \mathcal{Q}^{pe}_{u} \! - \! \zeta \! \cdot \! \sum_{\boldsymbol{\mathcal{R}}} \! \mathcal{Q}^{la}_{r} \quad \textit{s.t.} \quad \text{C1 - C18} \tag{OF} \\
& \mathcal{X}_r = \prod_{\phi \in \{pr,in,po\}} ( \sum_{F^{\phi}_{s_r},\mathcal{T}^{\phi}_r} \mathcal{X}^{\phi}_{r,f,t} ) \qquad \qquad \qquad \qquad \quad \forall \; r \in \boldsymbol{\mathcal{R}} \tag{C1}
\end{align}
\normalsize

 \subsection{Definition}
The definition constraints govern activation rules, logical dependencies among functions, and their placement across the network. Function placements are binary variables $\mathcal{Y}^{pr}_{f,n,t}, \mathcal{Y}^{po}_{f,n,t}$ indicating that function $f$ is deployed on node $n$ at time $t$, and $\mathcal{Y}^{in}_{r,n,t}$ indicates that request $r$'s inference task is served by node $n$ at $t$.
For notational uniformity, the same stage-indexed convention is used for placement and computation variables, with the inference stage treated as a singleton component.
Constraint C2 ensures each request $r \in \mathcal{R}$ activates $\pi^{pr}_{s_r}$ pre-processing function $f \in \mathcal{F}^{pr}_{s_r}$ at entry time, and $\pi^{po}_{s_r}$ post-processing function $f \in \mathcal{F}^{po}_{s_r}$ at the terminal time. Constraint C3 binds inference execution variables $\mathcal{X}^{in}_{r,t}$ to deployment variables $\mathcal{Y}^{in}_{r,n,t}$, ensuring scheduled inference is served by a node. Finally, Constraint C4 guarantees that if a function $f$ is demanded at time $t$, it should be deployed on at least one node. 

\footnotesize
\begin{align}\label{definitions}
& \sum_{\boldsymbol{\mathcal{F}}^{\phi}_{s_r}}\mathcal{X}^{\phi}_{r,f, \mathcal{T}^{\phi}_r} = \pi^{\phi}_{s_r} \qquad \qquad \qquad \qquad \qquad \forall \; r, \phi \in \boldsymbol{\mathcal{R}}, \; \{pr, po\} \tag{C2} \\
& \sum_{\boldsymbol{\mathcal{N}}} \mathcal{Y}^{in}_{r,n,t} > \sum_{\boldsymbol{\mathcal{R}}} \mathcal{X}^{in}_{r,t} \qquad \qquad \qquad \qquad \qquad \qquad \forall \; r,t \in \boldsymbol{\mathcal{R}}, \Delta_r \tag{C3} \\
& \sum_{\boldsymbol{\mathcal{N}}} \mathcal{Y}^{\phi}_{f,n,t} > \frac{ \sum_{\boldsymbol{\mathcal{R}}} \mathcal{X}^{\phi}_{r,f,t} }{\mathcal{R}} \qquad \quad \forall f,t, \phi \in \bigcup(\boldsymbol{\mathcal{F}}^{\phi}_s), \Delta_r, \; \{pr, po\} \tag{C4}
\end{align}
\normalsize

\subsection{Multi-time-frame Trajectory}
To enable mobility-aware orchestration of aerial nodes, these constraints govern \ac{UAV} spatial behavior and user interactions. \ac{UAV} positions are captured by $\mathcal{S}_{n,a,t}$, which equals 1 if \ac{UAV} $n$ occupies an area $a$ at time $t$, while edge nodes remain static but area-aware. Constraint C5 enforces location exclusivity, requiring each \ac{UAV} to occupy exactly one area per time slot. Constraint C6 introduces adjacency-based transitions, allowing a \ac{UAV} to move to an area $a$ at time $t$ only if it resided in an adjacent area $a'$ at the preceding time $t-1$.
Constraint C7 ensures a consistent user-node association $\mathcal{B}_{u,n,t}$ by restricting each user $u$ to at most one binned node per time frame, and activating it only when the user and node share the same area.

\footnotesize
\begin{align}\label{uav_trajectory}
    & \sum_{\boldsymbol{\mathcal{A}}} \mathcal{S}_{n,a,t} 
    = 1 \qquad \qquad \qquad \qquad \qquad \qquad \qquad \quad \forall \; n, t \in \boldsymbol{\mathcal{N}}, \boldsymbol{\mathcal{T}} \tag{C5} \\
    & \mathcal{S}_{n,a,t} \leq \sum_{a' \in \boldsymbol{\mathcal{A}}} \mathcal{A}_{a,a'} \cdot \mathcal{S}_{n,a',t-1}  \qquad \qquad \quad \forall \; n,a,t \in \boldsymbol{\mathcal{N}}, \boldsymbol{\mathcal{A}}, \boldsymbol{\mathcal{T}} \tag{C6} \\
    & \sum_{\boldsymbol{\mathcal{N}},\boldsymbol{\mathcal{A}}} \mathcal{B}_{u_r,n,t} \cdot \mathcal{S}_{n,a,t} \cdot \mathcal{I}_{u_r,a,t} \leq 1
    \qquad \qquad \qquad \forall \; r,t \in \boldsymbol{\mathcal{R}}, \Delta_r \tag{C7}
\end{align}
\normalsize

\subsection{Dynamic Network Graph}
These constraints govern the dynamic construction of time-varying network topology under mobility. Constraint C8 defines $\mathcal{L}_{n,n',t}$, a binary variable indicating the existence of a link between the nodes $n$ and $n'$ at time $t$, enabled only when nodes occupy adjacent or overlapping areas, thus reflecting nodes' coverage limits. Based on this, the feasible link set $\boldsymbol{\mathcal{L}}_t$ is constructed (C9), while the candidate path set $\boldsymbol{\mathcal{P}}_t$ is derived as sequences of active links (C10). Constraint C11 then enforces routing consistency by activating $\mathcal{J}_{p,l,t}$ only if link $l$ is active in path $p$ at $t$.

\footnotesize
\begin{align}\label{link_calculation}
& \mathcal{L}_{n,n',t} \leq \!\!\!\! \sum_{a,a' \in \boldsymbol{\mathcal{A}}} \!\! \mathcal{A}_{a,a'} \cdot \mathcal{S}_{n,a,t} \cdot \mathcal{S}_{n',a',t} \qquad \forall \; n,n',t \in \boldsymbol{\mathcal{N}}, \boldsymbol{\mathcal{N}}, \boldsymbol{\mathcal{T}} \tag{C8}  \\
& \boldsymbol{\mathcal{L}}_t \triangleq \left\{ (n, n') \in \boldsymbol{\mathcal{N}} \times \boldsymbol{\mathcal{N}} \;~\middle|~\; \mathcal{L}_{n,n',t} = 1 \right\} \qquad \qquad \quad \forall \; t \in \boldsymbol{\mathcal{T}} \tag{C9} \\
& \boldsymbol{\mathcal{P}}_t \triangleq \left\{ p = (\mathcal{H}_{p,t} \;, \mathcal{T}_{p,t}) ~\middle|~ p \subset \boldsymbol{\mathcal{L}}_t \right\} \qquad \qquad \qquad \qquad \forall \; t \in \boldsymbol{\mathcal{T}} \tag{C10} \\
& \mathcal{J}_{p,l,t} \leq \sum_{\boldsymbol{\mathcal{N}}, \boldsymbol{\mathcal{N}}} \mathcal{L}_{n,n',t} \qquad \qquad \qquad \qquad \quad \forall \; p,l,t \in \boldsymbol{\mathcal{P}}_t, \boldsymbol{\mathcal{L}}_t, \boldsymbol{\mathcal{T}} \tag{C11}
\end{align}
\normalsize

\subsection{Path Selection}
These constraints structure data flow across service stages, ensuring that requests are routed on feasible, coherent paths in dynamic topologies induced by \ac{UAV} mobility.
Routing is enforced across service stages. For each request $r$, time frame $t$, and stage $\phi$, one feasible path is selected between the active source and destination endpoint (C12). 
Here, $\eta^{h,\phi}_{r,n,t}$ and $\eta^{t,\phi}_{r,n,t}$ denote the active head and tail nodes of the stage $\phi$'s route for request $r$ at time $t$: for $\phi=pr$, they correspond to the user's PoA node and the selected pre-processing node; for $\phi=in$, to the previously selected pre-processing node and the selected inference node; and for $\phi=po$, to the selected inference node and the selected post-processing node.
In this way, the path is created whenever two consecutive functions or inference nodes are active at frame $t$, which adapts to time-varying connectivity, resource availability, and user proximity.

\footnotesize
\begin{align}
\label{path_selection}
\!\!\!\!\!\!\!\sum_{\substack{
\boldsymbol{\mathcal{P}}_t,\;
\boldsymbol{\mathcal{F}}^{\phi}_{s_r}  \\
\mathcal{H}_{p,t} = \eta^{h,\phi}_{r,n,t}, \\
\mathcal{T}_{p,t} = \eta^{t,\phi}_{r,n,t}
}} \!\!\!\!\!\!\!\!\!\!
\mathcal{R}^{tr}_{r,p,t} \mathbbm{1}(\mathcal{Y}^{\phi}_{f,n,t} \!==\! 1) \!=\! 1
\quad \forall r,t, \phi \in \boldsymbol{\mathcal{R}}, \mathcal{T}^{\phi}_r, \{pr,in,po\}
\tag{C12}
\end{align}
\normalsize

\subsection{QoS Constraints}
These constraints ensure service delivery with sufficient quality. Constraint C13 models \ac{E2E} latency $\mathcal{Q}^{la}_{r}$ as the cumulative sum of link latencies, each given by the ratio of the request's maximum packet size $\widecheck{\mathcal{Z}}_{r,t}$ to link capacity $\widehat{\mathcal{L}}_{l}$. The latency should follow the request's predefined tolerance $\widecheck{\mathcal{Q}}^{la}_{r}$. Meanwhile, output quality $\mathcal{Q}^{ou}_r$ is captured as the cumulative fidelity contributed by the inference nodes serving $r$, while it should meet or exceed a required threshold $\widecheck{\mathcal{Q}}^{ou}_r$ (C14). To foster personalization quality, Constraint C15 introduces a recency-weighted history variable $\mathcal{H}^{pa}_{u,n,t}$, quantifying the degree to which a node $n$ has historically served user $u$ up to time $t$, modulated by a decay factor $\delta$. Building upon this, Constraint C16 defines $\mathcal{Q}^{pe}_u$ as the sum of such affinities across inference tasks, rewarding continuity-based responses.

\footnotesize
\begin{align}\label{qos_constraints}
    & \mathcal{Q}^{la}_{r} \triangleq \sum_{\boldsymbol{\mathcal{P}}_t,\boldsymbol{\mathcal{L}}_t, \boldsymbol{\mathcal{T}}} \mathcal{R}^{tr}_{r,p,t} \cdot \mathcal{J}_{p,l,t} \cdot \left( \widecheck{\mathcal{Z}}_{r,t} / \widehat{\mathcal{L}}_{l} \right) \leq \widecheck{\mathcal{Q}}^{la}_{r} \qquad \forall \; r \in \boldsymbol{\mathcal{R}} \tag{C13} \\
    & \mathcal{Q}^{ou}_{r} \triangleq \sum_{\boldsymbol{\mathcal{T}}, \boldsymbol{\mathcal{N}}} \mathcal{Y}^{in}_{r,n,t} \cdot \overline{Q}_n \geq \widecheck{\mathcal{Q}}^{ou}_{r} \qquad \qquad \qquad \qquad \quad \forall \; r \in \boldsymbol{\mathcal{R}} \tag{C14} \\
    & \mathcal{H}^{pa}_{u,n,t} = \sum^{t-1}_{\mathfrak{T}=0} \delta^{t-\mathfrak{T}} \sum_{\boldsymbol{\mathcal{R}}_u} \mathcal{Y}^{in}_{r,n,\mathfrak{T}} \qquad \qquad \qquad \forall \; u,n,t \in \boldsymbol{\mathcal{U}}, \boldsymbol{\mathcal{N}}, \boldsymbol{\mathcal{T}} \tag{C15} \\
    & \mathcal{Q}^{pe}_{u} \triangleq \sum_{\boldsymbol{\mathcal{T}}, \boldsymbol{\mathcal{N}}} \mathcal{H}^{pa}_{u,n,t} \sum_{\boldsymbol{\mathcal{R}}_u} \mathcal{Y}^{in}_{r,n,t} \qquad \qquad \qquad \qquad \qquad \forall \; u \in \boldsymbol{\mathcal{U}} \tag{C16}
\end{align}
\normalsize

\subsection{Capacity Constraints}
To ensure feasible service provisioning aligned with physical infrastructure limitations in the \ac{6G} aerial-terrestrial networks, capacity constraints bound both compute and communication resources. Computing load is limited by aggregating processing demands from all active pre-, inference, and post-processing tasks, requiring total utilization not exceeding the node's capacity $\widehat{\mathcal{C}}_n$ (C17). Communication is restricted by ensuring that cumulative bandwidth demands $\widecheck{\mathcal{L}}_r$ from all routed requests using the link $l$ do not surpass its capacity $\widehat{\mathcal{L}}_l$, thereby preventing congestion (C18).

\footnotesize
\begin{align}\label{capacity_constraints}
    &\sum_{\boldsymbol{\mathcal{R}},\boldsymbol{\mathcal{F}}^{\phi}_{s_r}} \!\!\! \mathcal{X}^{\phi}_{r,f,t} \! \cdot \mathcal{Y}^{\phi}_{f,n,t} \! \cdot \widecheck{\mathcal{C}}^{\phi}_{r,f} \leq \widehat{\mathcal{C}}_n \quad \forall \; n, t, \phi \in \boldsymbol{\mathcal{N}}, \boldsymbol{\mathcal{T}}, \{pr, in, po\} \tag{C17} \\ 
    & \sum_{\boldsymbol{\mathcal{R}}, \boldsymbol{\mathcal{P}}_t} \mathcal{R}^{tr}_{r,p,t} \cdot \mathcal{J}_{p,l,t} \cdot \widecheck{\mathcal{L}}_r \leq \widehat{\mathcal{L}}_{l} \qquad \qquad \qquad \qquad \forall \; l, t \in \boldsymbol{\mathcal{L}}_t, \boldsymbol{\mathcal{T}} \tag{C18}
\end{align}
\normalsize

\section{Proposed Method}\label{sec:method}
The problem defined in Section \ref{sec:problem} is NP-hard \cite{deeplearningbasedservice}. Hence, finding the solution of (OF) becomes computationally intractable in large-scale instances despite unrealistic assumptions that all system knowledge is available. To address this, we propose the \ac{HyPE} framework, which integrates predictive modeling, learning-augmented decisions, and heuristics for scalable real-time service provisioning. \ac{HyPE} operates in three phases: (i) \ac{MAP}, which forecasts user request patterns to mitigate imperfect knowledge arising from user mobility; (ii) \ac{LEAD}, which leverages \acp{LLM} to jointly plan \ac{UAV} trajectories and inference node assignments based on historical context and quality requirements; and (iii) \ac{SET}, which applies \ac{PGA} for pre/post-function placement and \ac{LBSP} for efficient route selection under dynamic topology. The \ac{E2E} process of \ac{HyPE} is shown in Algorithm \ref{alg:HyPE}.

\textbf{MAP}: In the first phase, \ac{HyPE} employs a \ac{DRL} predictor to address the challenges of forecasting user mobility and anticipating service demands in highly dynamic \ac{6G} environments. This design builds on prior \ac{DRL}-based prediction methods for mobile service environments \cite{globecom2023}. Unlike offline learning models, which cannot adapt to rapidly changing traffic patterns, we employ online \ac{DRL} to continuously capture the spatio-temporal evolution of user behavior. Specifically, each \ac{PoA} hosts a \ac{D3QL} agent with a \ac{LSTM}-\ac{CNN} architecture to process historical mobility and request traces. The \ac{LSTM} captures temporal dependencies in user movement and service arrivals, while the \ac{CNN} extracts local spatial correlations across neighboring areas. Based on these observations, the agent outputs the probability distribution that user $u_r$ will appear in area $a$ and generate request $r$ in the next time frame. Unlike using prediction only for mobility estimation, \ac{MAP} is extended to jointly anticipate both user location and service intent, and to produce a prioritized set of likely future requests, reflecting their probability of occurrence. The agent's reward function is defined by prediction accuracy, incentivizing reliable mobility awareness while ensuring responsiveness to evolving user contexts. This “\ac{MAP}” does not output final \ac{UAV} placement or inference decisions; rather, it provides probabilistic forecasts of future user locations and anticipated requests that guide subsequent phases in proactive service provisioning.


\textbf{LEAD}: Following user mobility and request predictions and to satisfy anticipated request demands, we leverage an \ac{LLM}-driven planner as an adaptive decision engine. In this setting, the \ac{LLM} is tasked with producing two critical outputs for each upcoming time frame: (i) the trajectory design $\mathcal{S}_{n,a,t+1}$, denoting the area assignment for each \ac{UAV}, and (ii) the inference assignment $\mathcal{Y}^{in}_{r,n,t+1}$, specifying which network node is responsible for processing the inference stage of each predicted request. The \ac{LEAD} phase is tasked to transform the raw predictions of the \ac{MAP} phase into actionable orchestration decisions. These decisions are made while jointly considering latency constraints (C13), fidelity requirements (C14), and personalization objectives (C16), thereby ensuring that the orchestration remains both feasible and user-centric.

\acp{LLM} offer a compelling alternative by enabling adaptive, in-context learning-based decision-making. First, \acp{LLM} can directly process inputs and align with the nature of personalized \ac{AI} service provisioning, where user intent and mobility patterns should be considered simultaneously. Second, unlike static optimization solvers, \acp{LLM} can be prompted with domain-specific rules to perform inference selection and trajectory planning without exhaustively enumerating all feasible configurations. This allows decisions to be generated in real-time, even under partial observability or incomplete mobility information. Third, the ability of \acp{LLM} to generalize from prior interactions and adapt to new mobility or service contexts introduces a form of transferable intelligence, thereby supporting more flexible decision generation across heterogeneous deployment scenarios \cite{wibisono2024unstructured}. 
Finally, the adoption of \acp{LLM} integrates naturally with the broader system vision of \ac{AI}-native \ac{6G} networks, where network control and service orchestration are increasingly entrusted to intelligent methods rather than rigid rule-based optimization.

\begin{figure}[t!]\centering
\includegraphics[width=2.5in]{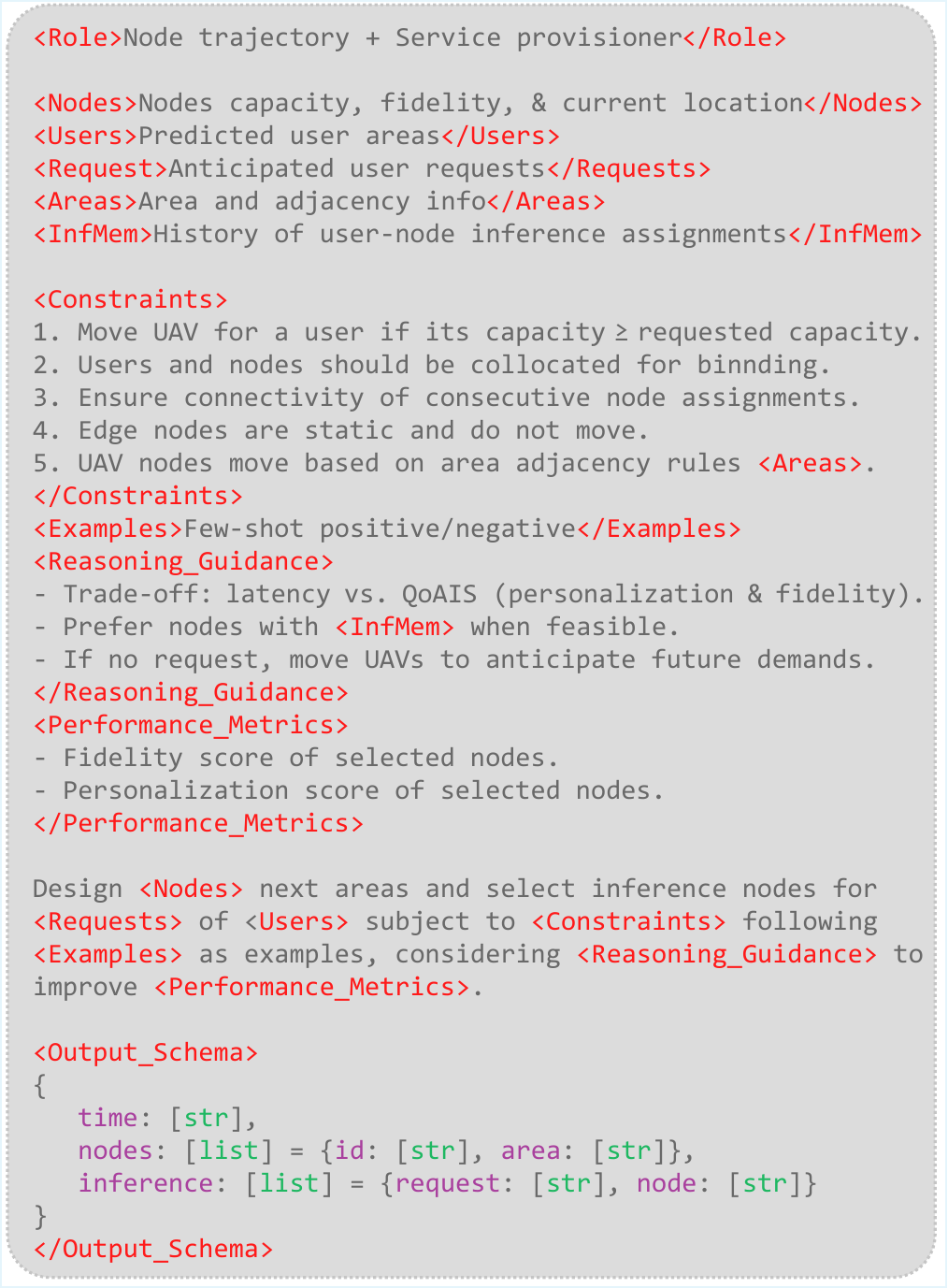}
\vspace{-5pt}
  \caption{Structured prompt design for LLM-based node trajectory and inference selection in the LEAD phase.}
    \label{fig:prompt_structure}
\end{figure}

In \acp{LLM}, a free-form text description is submitted to the language model via a prompt, and it is instructed to generate a structured output. To effectively harness the reasoning capacity of \acp{LLM}, we design a structured prompting strategy composed of three complementary mechanisms, depicted in Figure \ref{fig:prompt_structure}.

\textit{Role Prompting}: Initially, we implement role prompting, a natural language processing technique that assigns specific roles, personas, or contexts to language models to elicit more specialized responses. This method leverages corresponding knowledge patterns, terminology, and reasoning approaches typical of the assigned role, which enhances contextual consistency and reduces ambiguity \cite{kong2023better}. The \ac{LLM} is designated as a “Node trajectory and service provisioner” for a grid-based aerial-terrestrial network, responsible for providing \ac{UAV} areas and inference nodes. By assigning this specialized role, the \ac{LLM} is guided to operate within the reasoning patterns and decision logic expected of a domain-specific planner, rather than generating unconstrained text. The prompt provides contextual blocks, including predicted user areas and requests, node capacities and qualities, prior inference-memory for personalization continuity \(\mathcal{M}^{in}_t\), and domain-specific constraints, with the instruction that its output should consist of available node blocks while prohibiting the inclusion of its own suggestions. Thus, this approach effectively narrows the search space and grounds the \ac{LLM}'s reasoning in the operational state, thereby improving its accuracy and response time.

\textit{Contrastive Few-Shot Learning}: To ensure adaptability to evolving user contexts, the prompt incorporates contrastive few-shot examples derived from before the task execution \cite{fewshort2025}. High-reward examples (i.e., past assignments that yielded low latency as well as high fidelity and personalization scores) and low-reward examples (i.e., assignments violating \ac{QoS} or leading to infeasible placements) are embedded into the prompt. This contrastive structure enables the \ac{LLM} to refine its reasoning by explicitly learning from both successes and failures, thereby improving generalization to unseen mobility patterns and heterogeneous service demands. Through this mechanism, the \ac{LLM} avoids repeating detrimental provisioning strategies and adapts more robustly to the stochasticity of user mobility. What is more, we employ contrastive learning to ensure that the \ac{LLM} remains responsive to dynamic contexts in which user requests evolve and require diverse contextual interpretations. Specifically, we retain the recorded inference selection outputs $\mathcal{Y}^{in}_{r,n,t}$ following the completion of request $r$. For subsequent user requests, we select \( \mathcal{K} \) examples from the highest-reward outputs, referred to as \( \mathbb{K}^+ \), and \( \mathcal{K} \) examples from the lowest-reward outputs, referred to as \( \mathbb{K}^- \). This strategy provides the \ac{LLM} with updated guidance on what to prioritize and what to avoid, ensuring its adaptability to shifting requirements, particularly those related to personalization.

\textit{Structure Enforcement}: To guarantee seamless integration of \ac{LLM} outputs into the orchestration pipeline, we enforce a structured response format. Specifically, the \ac{LLM} is instructed to provide node area and inference node assignments in a machine-readable schema, validated automatically using the Pydantic library \cite{pydantic}. 
The schema fields directly map to the optimization variables: nodes candidates $\mathcal{S}_{n,a,t+1}$ and inference candidates $\mathcal{Y}^{in}_{r,n,t+1}$, while validated outputs are the only decisions passed to \ac{SET} for $\mathcal{B}_{u,n,t+1}$, $\mathcal{Y}^{pr,po}_{f,n,t+1}$, and $\mathcal{R}^{tr}_{r,p,t+1}$. 
Furthermore, the structured format allows direct validation against feasibility constraints: \ac{UAV} moves are checked against adjacency rules (C6), inference assignments are validated against node capacity constraints (C17), and the network graph is updated according to link feasibility (C9-C10). Any infeasible decisions are discarded, and the \ac{LLM} is iteratively guided toward valid outputs.

The \ac{LEAD} phase algorithm is shown in Algorithm \ref{alg:HyPE}, steps 2-8. Consider a network with one \ac{UAV}, one edge node, and two users $u_1$ and $u_2$. At time $t$, the \ac{MAP} phase predicts that $u_1$ will move to area $a_2$ and request the speech-to-text service, while $u_2$ will remain in area $a_1$ and request image captioning. The \ac{UAV} is currently in area $a_1$, and the edge node is static in $a_3$. The \ac{LLM} prompt contains: (1) $<\!\!Users\!\!>$: predicted areas and $<\!\!Requests\!\!>$ ($u_1$ in $a_2$, $u_2$ in $a_1$); (2) $<\!\!Nodes\!\!>$: \ac{UAV} (capacity 100 \ac{GFLOPS}, fidelity 0.6), edge (capacity 400 \ac{GFLOPS}, fidelity 0.8); (3) $<\!\!InfMem\!\!>$: $u_1$ previously served by edge, $u_2$ by \ac{UAV}; (4) $<\!\!Reasoning\_Guidance\!\!>$: maximum tolerable latency 80ms, minimum required fidelity 0.7. The \ac{LLM}, acting as the \ac{UAV} trajectory and service provisioner, produces: (1) \ac{UAV} trajectory: candidate move of \ac{UAV} from $a_1$ to $a_2$ 
($\mathcal{S}_{uav,a_2,t+1} = 1$); (2) Inference assignment: candidate assignment of $u_1$'s inference on the edge node ($\mathcal{Y}^{in}_{r_{u_1},edge,t+1} = 1$) to meet fidelity $>$ 0.7, and keep $u_2$'s inference on the \ac{UAV} ($\mathcal{Y}^{in}_{r_{u_2},uav,t+1} = 1$) to minimize latency and increase personalization. After validation, these assignments are feasible: \ac{UAV} mobility satisfies adjacency (C6), inference node assignments respect node capacities (C17), and the updated graph reflects new links (C9-C10). By validating outputs against network constraints, the framework ensures that \ac{UAV} trajectories and node selections are not only context-aware but also operationally feasible. Hence, LEAD is not treated as an exact solver; it serves as a structured policy prior that proposes high-utility candidate decisions, which are then completed by deterministic constraint validation and \ac{SET}. Thus, the validated outputs feed into the subsequent \ac{SET} phase, where pre/post-processing placement and routing decisions are finalized to complete the orchestration cycle.

\textbf{SET}: This phase integrates (i) \ac{MAP}'s predicted user areas and requests, (ii) \ac{LEAD}'s inference-node and trajectory decisions, and (iii) residual capacities to finalize user-node bindings, placing pre-/post-processing functions, and selecting routing paths, as detailed in Algorithm \ref{alg:HyPE}, steps 9-25. First, binned node ($\mathcal{B}_{u,n,t+1}$) is selected for each active user $u$ at time $t\!+\!1$, as required by C7. Let $\boldsymbol{\mathcal{N}}^{bi}_{u_r,t+1}\!$ be the candidate collocated nodes for $u$, as defined in Eq.~(\ref{eq:candidate_binding_nodes}). The binding is selected by minimizing latency to $n'$ from $u$'s \ac{PoA} ($\mathrm{L}_{u\rightarrow n'}$) while incorporating historical affinity $\widehat{\mathcal{H}}_{u,n,t+1}$ (normalized history value from $\mathcal{H}^{pa}_{u,n,t+1}$), weighted by $\kappa$ to balance recency with latency, as shown in Eq.~(\ref{eq:set_binding}). 

\footnotesize
\vspace{-6pt}
\begin{equation}
\boldsymbol{\mathcal{N}}^{bi}_{u,t+1} = \{ n \in \mathcal{N} \;|\; \exists a: \mathcal{I}_{u,a,t+1}=1 \ \land\ \mathcal{S}_{n,a,t+1}=1 \},
\label{eq:candidate_binding_nodes}
\end{equation}
\vspace{-7pt}
\begin{equation}
\mathcal{B}_{u,n,t+1} \!\!=\!\!
\begin{cases}
1, & \text{if } n \!=\! \arg\!\!\!\!\!\!\!\!\!\min\limits_{n'\in \boldsymbol{\mathcal{N}}^{bi}_{u,t+1}} \!\!\!\! \Big( \mathrm{L}_{u\rightarrow n'} \! + \! \kappa\,(1-\widehat{\mathcal{H}}_{u,n',t}) \Big). \\
0, & \text{otherwise}.
\end{cases}
\label{eq:set_binding}
\end{equation}
\vspace{-6pt}
\normalsize

\footnotesize
\begin{algorithm}[t]
\caption{HyPE service provisioning}
\label{alg:HyPE}
\KwIn{$\boldsymbol{\mathcal{N}}_t$ (state, capacity, output fidelity)}

\KwOut{$\mathcal{S}_{n,a,t+1}$,$\mathcal{Y}^{in}_{r,n,t+1}$,$\mathcal{Y}^{pr,po}_{f,n,t+1}$,$\mathcal{B}_{u,n,t+1},\!\mathcal{R}^{tr}_{r,p,t+1}$}

Use DRL method to predict $\mathcal{I}_{u,a,t+1}$, $\boldsymbol{\mathcal{R}}_{t+1}$

Create $\mathcal{M}^{in}_t$ \& block structured prompt with examples \\



Query LLM \& receive candidates $\{\mathcal{S}_{n,a,t+1}, \, \mathcal{Y}^{in}_{r,n,t+1} \}$


\ForEach{UAV $n \in \boldsymbol{\mathcal{N}}_t$}{
    Validate adjacency constraint (C6)
}
\ForEach{$r \in \boldsymbol{\mathcal{R}}_{t+1}$}{
    Validate node (C17) \& fidelity/latency (C13-C14)
}

Update network graph $\mathcal{G}_{t+1}$ \& $\mathcal{M}^{in}_{t+1}$

\ForEach{$r \in \boldsymbol{\mathcal{R}}_{t+1}$}{
  Compute candidate set $\boldsymbol{\mathcal{N}}^{bi}_{u,t+1}$ (Eq.~(\ref{eq:candidate_binding_nodes})) \\
  Compute $\mathrm{L}_{u\rightarrow n}$ \& $\mathcal{B}_{u,n^\star,t+1}$ based on Eq.~(\ref{eq:set_binding})
}

Sort $\!\mathcal{F}_{t+1}\!$ by latency tightness $\sum_{\boldsymbol{\mathcal{R}}_f}\!\! r | \boldsymbol{\mathcal{R}}_f \!\!\leftarrow\!$ f's requests \\
\ForEach{$f \in \mathcal{F}_{t+1}$}{
  Compute candidate nodes $\boldsymbol{\mathcal{N}}^{fe}_f$ that meet Eq.~(\ref{eq:set_pg_capacity}) \\
  \If{$\boldsymbol{\mathcal{N}}^{fe}_f = \varnothing$}{
    Spillover\_replication (partition $\boldsymbol{\mathcal{R}}_f$)
  }
  $n^\star = \arg\!\min_{n\in\boldsymbol{\mathcal{N}}^{fe}_f} \sum_{\boldsymbol{\mathcal{R}}_f} \mathrm{L}_{u_r \rightarrow n}$ \\
  $\mathcal{Y}^{pr,po}_{f,n^\star,t+1} \leftarrow 1$ \& Update $\widehat{\mathcal{C}}_{n^\star}$ (Eq.~(\ref{eq:update_node_capacities}))
}

\ForEach{$r \in \boldsymbol{\mathcal{R}}_{t+1}$}{
  Determine $r$'s stage \& Find previous node \\
  \ForEach{$l \in \boldsymbol{\mathcal{L}}_t$}{
    Compute $w_{l,r,t+1}$  based on Eq.~(\ref{eq:set_lbsp_weight}) \\
    \While{found}{
        Find $p^\star$ via Eq.~(\ref{eq:set_lbsp_capacity})
    }
    $\mathcal{R}^{tr}_{r,p^\star,t+1} \leftarrow 1$ \& Update $u_{l,t+1} \; \forall \; l \in p^\star$
  }
}

\end{algorithm}
\normalsize

Next, the \ac{PGA} method is used for function placement ($\mathcal{Y}^{pr,po}_{f,n,t+1}$) that assigns each required function $f \in \mathcal{F}_{t+1}$ to the node that minimizes aggregated latency for its requesting users $~\boldsymbol{\mathcal{R}}_f$. Functions are prioritized by latency tightness ($\sum_{\boldsymbol{\mathcal{R}}_f} r$). For each $f \in \mathcal{F}_{t+1}$, we define the candidate nodes set $\mathcal{N}^{fe}_f$ that satisfy Eq.~(\ref{eq:set_pg_capacity}), where $p \cdot \widehat{\mathcal{C}}_n$ denotes node $n$'s residual compute capacity. If no candidate node exists ($\mathcal{N}^{fe}_f\!=\!\varnothing$), $f$'s requests ($\mathcal{R}_f$) are partitioned across replicas on multiple nodes based on spillover replication, which assigns replicas greedily until the subset of nodes that satisfies (Eq.~\ref{eq:set_pg_capacity}). In other words, we find two or more nodes to satisfy the function $f$, and each node serves a subset of requests. Afterward, for each feasible node, the total latency is $\sum_{\boldsymbol{\mathcal{R}}_f} \mathrm{L}_{u_r \rightarrow n}$, where $\mathrm{L}_{u_r \rightarrow n}\!$ is computed over feasible paths in $t+1$. The placement node $n^\star = \arg\!\min_{n\in\boldsymbol{\mathcal{N}}^{fe}_f} \sum_{\boldsymbol{\mathcal{R}}_f} \mathrm{L}_{u_r \rightarrow n}$ is then selected, and capacities are updated via Eq.~(\ref{eq:update_node_capacities}). The process repeats until all $\mathcal{F}_{t+1}$ functions are placed or resources are exhausted.

\footnotesize
\vspace{-1pt}
\begin{equation}
\sum_{\boldsymbol{\mathcal{R}}_f} (\widecheck{\mathcal{C}}^{pr}_{r,f} + \widecheck{\mathcal{C}}^{po}_{r,f}) \; \le \; p \cdot \widehat{\mathcal{C}}_n,
\label{eq:set_pg_capacity}
\end{equation}
\vspace{-6pt}
\begin{equation}
p \cdot \widehat{\mathcal{C}}_{n^\star} = p \cdot \widehat{\mathcal{C}}_{n^\star} - (\sum_{\boldsymbol{\mathcal{R}}_{f_{pr}}} \widecheck{\mathcal{C}}^{pr}_{r,f} + \sum_{\boldsymbol{\mathcal{R}}_{f_{po}}} \widecheck{\mathcal{C}}^{po}_{r,f}).
\label{eq:update_node_capacities}
\end{equation}
\vspace{-6pt}
\normalsize

After binding and placement, routing is performed on the updated graph $\mathcal{G}_{t+1}$ using \ac{LBSP}, which selects low-latency paths while ensuring link capacity feasibility (C18). For each request $r$ and transfer step (user$\to$pre, pre$\to$inference, inference$\to$post), a feasible path $\mathcal{R}^{tr}_{r,p,t+1}$ is selected. Each request imposes latency based on $\widecheck{\mathcal{Z}}_{r,t}$ on link $l$, while load $u_{l,t+1}$ derived from \ac{LEAD} or earlier \ac{SET} steps. For each request, the per-link weight $w_{l,r,t+1}$ is defined via Eq.~(\ref{eq:set_lbsp_weight}), where $\gamma$ tunes congestion avoidance (practical tuning knob). The final path $p^\star$ minimizes latency while satisfying capacity constraints via Eq.~(\ref{eq:set_lbsp_capacity}). If the shortest path is infeasible, \ac{LBSP} iteratively tests alternatives; if none exist, the request is delayed (if $\Delta_r$ allows) or rejected.

\footnotesize
\vspace{-4pt}
\begin{equation}
    w_{l,r,t+1} = \frac{\widecheck{\mathcal{Z}}_{r,t}}{\widehat{\mathcal{L}}_{l}} + \gamma \cdot \frac{u_{l,t+1}}{{\widehat{\mathcal{L}}_{l}}},
\label{eq:set_lbsp_weight}
\end{equation}
\vspace{-8pt}
\begin{align}
    & \! p^\star \!=\! \arg\!\!\!\!\!\min_{p \in \boldsymbol{\mathcal{P}}_{t\!+\!1}} \sum_{l \in p} w_{l,r,t+1} \; \text{s.t.} \sum_{\tilde r, \tilde p \ni l} \!\! \mathcal{R}^{tr}_{\tilde r,\tilde p,t+1} \! \cdot \! \widecheck{\mathcal{L}}_{\tilde{r}} \!+\! \widecheck{\mathcal{L}}_{r} \!\! \le \!\! {\widehat{\mathcal{L}}_{l}} 
\; \; \forall l \in p^\star.
\label{eq:set_lbsp_capacity}
\end{align}
\normalsize

\section{Performance Evaluation}\label{sec:results}
Our simulation environment is designed to reproduce the computational and networking characteristics of \ac{6G} networks. Table~\ref{tab:sim-parameters} lists the key parameters used in the simulation environments. To capture spatial dynamics, we map real-world user mobility traces from Zenodo \cite{yjmobilityDataset} to our grid-based system that reflects real-world trajectories. Complementing mobility, requests are derived from the MMMU-Pro dataset \cite{MMMUdataset}, which provides diverse \ac{AI} tasks. By probabilistically associating tasks to users over time, we reproduce the heterogeneous, unpredictable workloads typical of \ac{6G} scenarios. 


We evaluate \ac{HyPE} against state-of-the-art, optimization-based, and random baselines. As strong comparators, we include \ac{AD-SAC} \cite{Yan2025}, a hybrid offloading/power-allocation method that minimizes latency under accuracy constraints via accuracy-aware inference offloading in \ac{UAV}-satellite networks, and \ac{JAAPD-D} \cite{Hu2024}, which jointly balances latency, acceptance rate, and resource orchestration with objectives overlapping ours. For a fair comparison, all methods are executed under the same constraints, such as mobility traces, request realizations, and latency/fidelity demands. To integrate them into our setting, \ac{AD-SAC} is used for \ac{UAV} trajectory and inference assignments under its original latency/fidelity design, while \ac{JAAPD-D} is adapted to the same aerial-terrestrial network as a demand-driven placement baseline. We also benchmark an optimization-based formulation (OF) solved via Gurobi and a random strategy that randomizes \ac{UAV} trajectories, function placements, and inference assignments. 

\footnotesize
\vspace{-2pt}
\begin{table}[t]
\centering
\caption{Simulation Parameters and System Configurations}
\vspace{-6pt}
\renewcommand{\arraystretch}{1}
\begin{tabular}{lcc}
\hline
\textbf{Parameter} & \textbf{Range / Value} & \textbf{Unit} \\
\hline
Service duration ($\mathcal{T}^{du}_s$) & 2-6 & Time Frames \\
User request bandwidth ($\widecheck{\mathcal{L}}_{r}$) & 100 – 1000 & Mbps \\
Pre-/Post-processing capacity ($\widecheck{\mathcal{C}}^{\phi}_{r,f}$) & 1 – 5 & GFLOPS \\
Inference capacity ($\widecheck{\mathcal{C}}^{in}_{r}$) & 30 – 100 & GFLOPS \\
Pre-/post-process step ($\pi^{\phi}_{s}$) & 1-2 & - \\
Personalization continuity factor ($\delta$) & 0.65 – 0.9 & - \\
Minimum output quality ($\check{\mathcal{Q}}^{ou}_r$) & 0.5 – 0.8 & - \\
Latency requirement ($\check{\mathcal{Q}}^{la}_r$) & 50 – 100 & ms \\
Max packet size ($\widecheck{\mathcal{Z}}_{r,t}$) & 256 – 2048 & Bytes \\
Node quality ($\overline{Q}_n$) & 0.6 - 1 & - \\
UAV compute capacity ($\widehat{\mathcal{C}}^{ca}_n $) & 80 – 150 & GFLOPS \\
Cloud/Edge compute capacity ($\widehat{\mathcal{C}}^{ca}_n $) & 500–1000 & GFLOPS \\
Link bandwidth capacity ($\widehat{\mathcal{L}}_l$) & 100 – 2000 & Mbps \\
MAP learning rate (D3QL) & $10^{-4}$ – $10^{-3}$ & - \\
MAP replay buffer size & 1k – 10k & transitions \\

LEAD LLM setup & \makecell{Gemini-2.5-Flash;\\ 1 API/frame; \\temp. 0.1} & - \\
LEAD LLM cost & \makecell{\$0.30,2.50/M} & in,out \\
LLM prompt length & 1k-4k & tokens \\
LLM few-shot examples ($\mathcal{K}$) & 1-3 & - \\

Break weight / LBSP bias ($\kappa, \gamma$) & 0.1 – 1.0 & - \\
Areas ($\boldsymbol{\mathcal{A}}$) / Edge-cloud/UAV nodes & 5x5 / 15 / 5 & - \\
\hline
\end{tabular}
\label{tab:sim-parameters}
\end{table}
\normalsize

\begin{figure}[t!]\centering
\includegraphics[width=2.2in]{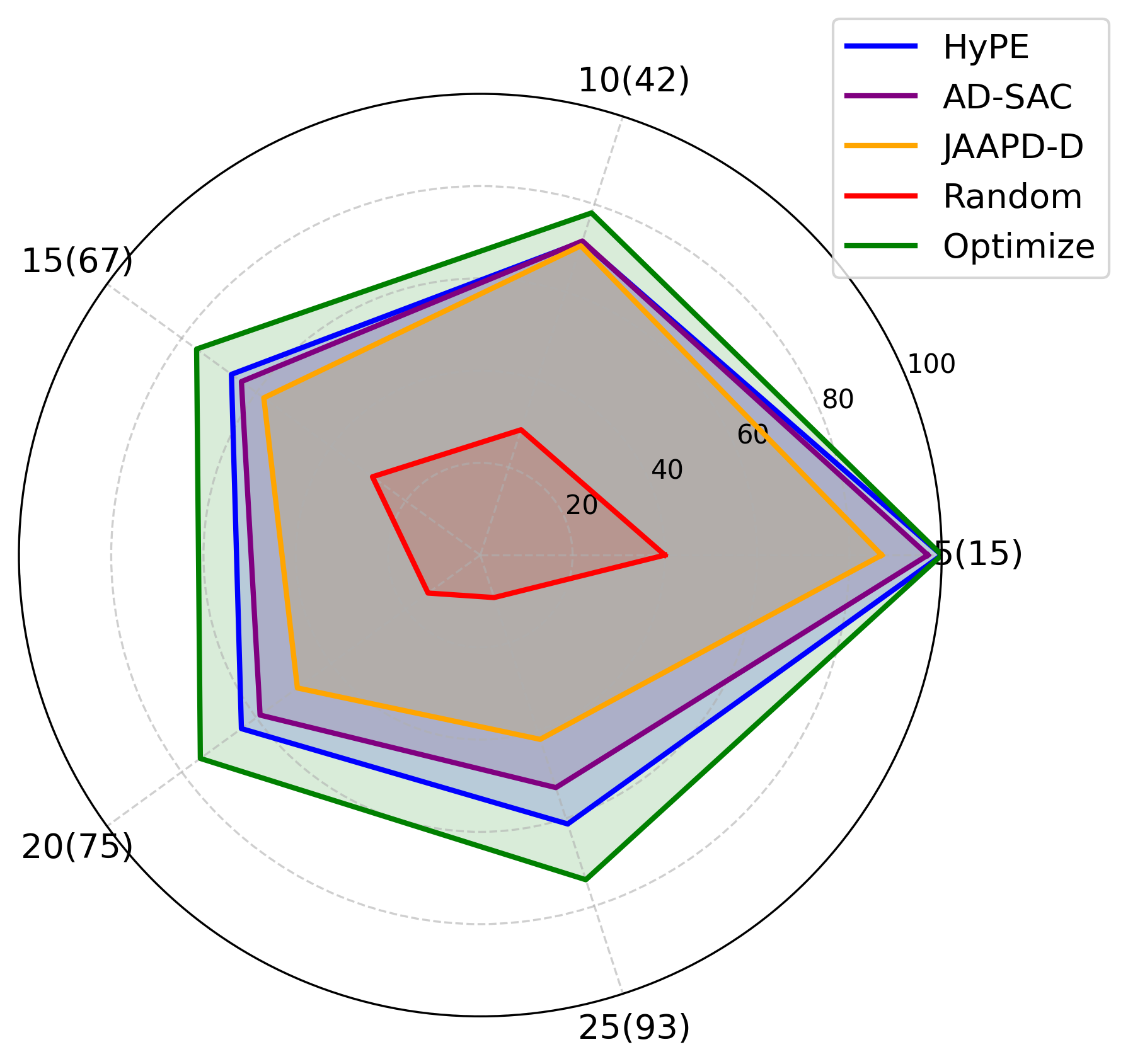}
\vspace{-6pt}
  \caption{Coverage comparison in terms of accepted requests as the number of users (requests) increases.}
    \label{figure:users_accepted}
\end{figure}

Our experiments assess how key 6G capabilities are sustained as the density grows \cite{ITU2160}.
The first scenario examines coverage, wherein parts of the area is served by fixed infrastructure, requiring \ac{UAV} repositioning for service delivery. As shown in Fig.~\ref{figure:users_accepted}, the Optimize oracle (full demand knowledge) achieves 100\% acceptance at light loads and 75\% at heavy loads, exposing \ac{UAV} unreachability due to adjacency-limited movements. The Random baseline performs worst as \acp{UAV} fail to track moving users, causing a sharp degradation. \ac{HyPE} matches optimal acceptance with 5 users and sustains 61\% under heavy load, a strong suboptimal result driven by resource saturation and \ac{MAP} prediction errors. \ac{AD-SAC} drops to 53\% because its agent optimizes \ac{UAV} placement without explicit allocation or per-request admission, inducing contention and latency violations. \ac{JAAPD-D} fares lower (42\%) as its demand-driven heuristic lacks foresight for proactive \ac{UAV} repositioning, leading to longer paths and reduced acceptance. By contrast, \ac{HyPE}’s history-driven \ac{LEAD} and prediction-aware planning better accommodate requests under load.

\begin{figure*}[t!]\centering
\includegraphics[width=6.8in]{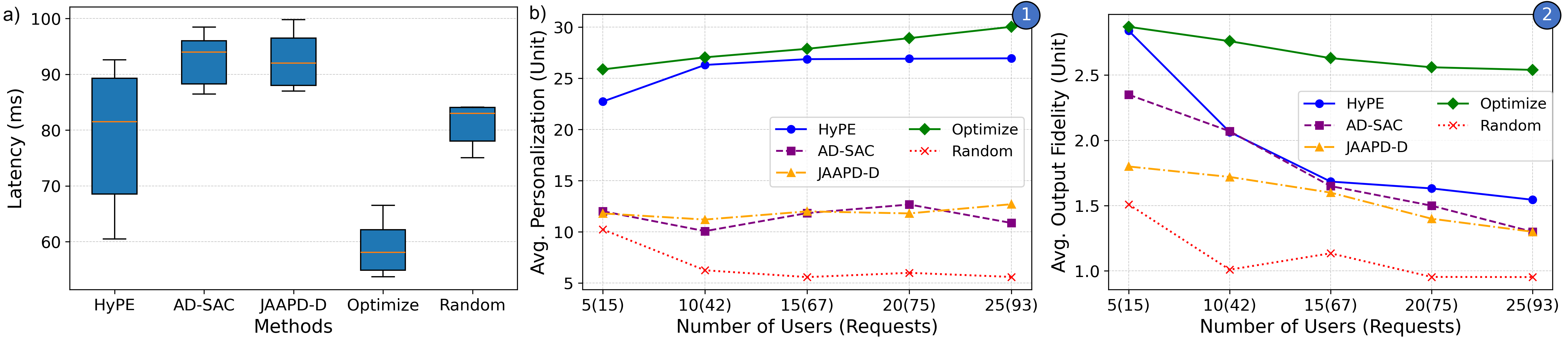}
\vspace{-5pt}
  \caption{Comparison of \ac{HyPE} with Optimize (OF), AD-SAC \cite{Yan2025}, JAAPD-D \cite{Hu2024}, and random methods in terms of a) latency, and b) AI-enabled capabilities (\ac{QoAIS} as Personalization and Fidelity) as the number of users/requests (connection density) expands. The lower bounds in (a) mainly correspond to light-load periods with fewer users and requests, reflecting underutilized resources and minimal contention.}
    \label{fig:users_final}
    \vspace{-4pt}
\end{figure*}

The second scenario examines how HyPE balances E2E latency and AI-enabled capabilities \cite{ITU2160} as demand increases, comparing Optimize (oracle), Random, \ac{AD-SAC}, \ac{JAAPD-D}, and \ac{HyPE}. Fig.~\ref{fig:users_final}.a reports latency distributions (min/max/median with means) for accepted requests: Optimize achieves the lowest latency via full demand knowledge and coordinated UAV control; Random shows deceptively lower mean latency than \ac{AD-SAC} and \ac{JAAPD-D} as it accepts few requests (mostly from fixed nodes), biasing results; \ac{HyPE} balances latency and \ac{QoAIS} in which at light load, LEAD and SET favor high-fidelity edge inference, slightly increasing latency (92 ms vs. 86 ms for \ac{AD-SAC} at 5 users), while at higher loads LBSP shortens paths and PGA shifts tasks to UAVs near users, reducing latency with modest, acceptable quality loss; \ac{AD-SAC} minimizes latency at low load via smaller UAV-hosted models but escalates under heavy load due to lack of bandwidth control, whereas LEAD's prediction-aware trajectories keep \ac{HyPE} within bounds; \ac{JAAPD-D} starts slightly lower through dynamic allocation but degrades without proactive UAV repositioning as density rises.

Fig.~\ref{fig:users_final}.b evaluates AI-enabled personalization (continuity of user experience) and per-request output fidelity. Optimize leads by jointly optimizing placement, UAV movement, and historical affinity, while Random fails on both metrics due to unguided mobility and placement. \ac{HyPE} achieves robust personalization and competitive fidelity by combining demand prediction with LEAD pre-positioning and PGA’s continuity-aware placement toward spatially proximal nodes; personalization increases by 19\% under load, while fidelity decreases (yet acceptable) by 45\% as tasks shift to UAVs (intended trade-off given \(\delta\)’s personalization weighting). Under heavier load, HyPE shifts a larger fraction of inference tasks to distilled models to preserve service continuity and timely response delivery when fixed edge resources become saturated. Rather than a collapse in service quality, this is a controlled shift in which HyPE maintains interactive latency and personalization while still retaining competitive fidelity. Compared to \ac{AD-SAC}, \ac{HyPE} attains higher \ac{QoAIS} by embedding personalization into reward and placement logic; \ac{AD-SAC} degrades on both due to smaller UAV LLMs and lack of affinity modeling. \ac{JAAPD-D} likewise underperforms \ac{HyPE} (personalization 12 vs. 26) because personalization is absent from its objectives.

\section{Conclusion}\label{sec:conclusion}
This paper addressed personalized AI services in UAV-assisted 6G networks under mobility and constrained edge resources. We introduced HyPE, a hybrid predictive-in-context-learning framework that integrates mobility-aware demand prediction, LLM-guided decision-making, and heuristic function placement and routing to maintain stringent latency and QoAIS guarantees. Empirically, HyPE delivers elastic provisioning across the three axes of 6G capabilities: trajectory-managed UAVs extend coverage to 91\% of the oracle while complementing fixed edges; accepted-request E2E latency remains within 6G-class interactive bounds; and distributed aerial-terrestrial inference attains 94\% of optimal output fidelity. By eschewing combinatorial enumeration, HyPE achieves polynomial per-frame complexity \( \mathcal{O}\!\left(\mathcal{T}(\mathcal{N}\mathcal{U} + \mathcal{N}\mathcal{U}\mathcal{F} + \mathcal{U}\mathcal{N}^2\log\mathcal{N})\right) \) trading worst-case optimality for real-time scalability.
Future work includes energy-aware UAV trajectory planning, cross-domain orchestration, and continual learning for long-term personalization. As HyPE leverages pre-trained LLMs, future deployment should account for inference costs on-device or in federated setups.

\section*{Acknowledgment}
The research work is supported in part by the Federal Ministry of Research, Technology, and Space (BMFTR), Germany, through the Project 6GEM+ under Grant 16KIS2411; by the European Union’s Horizon Europe research and innovation programme under the 6G-Path project (Grant No. 101139172); and the Research Council of Finland 6G Flagship Programme under Grant No. 369116.

\bibliographystyle{IEEEtran}
\bibliography{References/conf_short, References/IEEEabrv, References/main}

\end{document}